\documentclass[12pt]{iopart}

\usepackage{amsfonts}

\usepackage[dvips]{graphicx}
\usepackage{iopams}
\usepackage{amsfonts}
\begin{document}

\title[Scaling and non-Abelian signature in FQH
quasiparticle tunneling amplitude]
{Scaling and non-Abelian signature in fractional quantum Hall
quasiparticle tunneling amplitude}

\author{Zi-Xiang Hu}
\address{Asia Pacific Center for Theoretical Physics,
Pohang, Gyeongbuk 790-784, Korea}
\address{Department of Electrical Engineering, Princeton University, 
Princeton, New Jersey 08544, USA}

\author{Ki Hoon Lee}
\address{Asia Pacific Center for Theoretical Physics,
Pohang, Gyeongbuk 790-784, Korea}
\address{Department of Physics, Pohang University of Science and
Technology, Pohang, Gyeongbuk 790-784, Korea}

\author{Edward H. Rezayi}
\address{Department of Physics, California State University
Los Angeles, Los Angeles, California 90032, USA}

\author{Xin Wan}
\address{Asia Pacific Center for Theoretical Physics,
Pohang, Gyeongbuk 790-784, Korea}
\address{Department of Physics, Pohang University of Science and
Technology, Pohang, Gyeongbuk 790-784, Korea}
\address{Zhejiang Institute of Modern Physics, Zhejiang
University, Hangzhou 310027, P.R. China}

\author{Kun Yang}
\address{National High Magnetic Field Laboratory and
Department of Physics, Florida State University, Tallahassee,
Florida 32306, USA}

\date{\today}

\begin{abstract}

  We study the scaling behavior in the tunneling amplitude when
  quasiparticles tunnel along a straight path between the two edges of a
  fractional quantum Hall annulus. Such scaling behavior originates
  from the propagation and tunneling of charged quasielectrons and
  quasiholes in an effective field analysis. In the limit when the
  annulus deforms continuously into a quasi-one-dimensional ring, we
  conjecture the exact functional form of the tunneling amplitude for
  several cases, which reproduces the numerical results in finite
  systems exactly. The results for Abelian quasiparticle tunneling is
  consistent with the scaling anaysis; this allows for the extraction of
  the conformal dimensions of the quasiparticles. We analyze the scaling
  behavior of both Abelian and non-Abelian quasiparticles in the
  Read-Rezayi ${\mathbb Z}_k$-parafermion states. Interestingly, the
  non-Abelian quasiparticle tunneling amplitudes exhibit nontrivial
  $k$-dependent corrections to the scaling exponent.

\end{abstract}

\maketitle

\section{Introduction}

Quasiparticle tunneling through narrow constrictions or point contacts that 
bring counter-propagating edges close could serve  as a powerful tool
to probe both the bulk topological order as well as edge
properties of fractional quantum Hall (FQH) liquids~\cite{kane92b}. In
particular, interference signatures from double point contact devices
may reveal the statistical properties of the quasiparticles that
tunnel through them~\cite{chamon97}, especially the non-Abelian
ones~\cite{stern06,bonderson06}. In recent interference experiments at
the $\nu=5/2$ FQH state~\cite{willett08,willett09}, Willett {\it et
  al.} found that quasiparticles with charge $e/4$ and $e/2$ both
contribute to the interference patterns and dominate in different
regimes, which was anticipated in earlier theoretical work\cite{wan08}. 
To have a complete understanding of these experiments, one
needs quantitative information on the relative importance of
quasiparticles with different charges. Motivated by this, four of the
authors and a co-worker~\cite{chen09} performed microscopic
calculations of the tunneling matrix elements of various types of
quasiparticles, for both the Abelian Laughlin state, and the
non-Abelian Moore-Read (MR) state. The focus of the previous work was
the dependence of these matrix elements on the tunneling distance: the
main result was that the ratio between tunneling matrix elements for
quasiparticles with different charges decays with tunneling distance
in a Gaussian form, which originates from their charge
difference. Such considerations and results are required for a
complete understanding of the non-Abelian interferometer~\cite{bishara09}.

On the other hand, the system size dependence of the tunneling matrix
elements is also an interesting issue. In microscopic studies, we
start from interacting electrons with fermionic statistics. With
proper choices of microscopic Hamiltonian, ground states with
nontrivial topological properties emerge, together with fractionally
charged quasiparticle excitations, which may obey either Abelian or
non-Abelian statistics. Naturally, in a calculation relevant to
quasiparticle tunneling amplitude we can read out the information of the
scaling dimension of the corresponding tunneling operator. In
particular, the finite system size cutoff in the numerical
calculations may introduce scaling behavior in the tunneling amplitude
with an exponent imprinted with the quasiparticle conformal dimension.

In the present paper we study the system size dependence of these
matrix elements in the Laughlin and the Moore-Read states. By
combining numerical calculations with effective field theory analysis, we
show that their size dependence takes power-law forms with exponents
related to the scaling dimensions of the corresponding quasiparticle
operators. Furthermore, in the limit when the annulus deforms
continuously into a quasi-one-dimensional ring, we conjecture the
precise functional forms of the size dependence, which is not only
consistent with the expected power-law form in the scaling limit, but
also verified to be true in finite-size systems (using the exact Jack
polynomial approach, rather than the Lanczos diagonalization with
controllable error), indicating their exactness. We also attempt to
extend the discussions to the Read-Rezayi states.

We review our model and earlier results in Sec.~\ref{sec:model}. In
Sec.~\ref{sec:fieldtheory} we formulate a scaling theory for the
tunneling amplitude of Abelian quasiparticles and compare it with
numerical scaling results. We then conjecture closed-form expressions
for the tunneling amplitude, from which we extract exact scaling
exponents in Sec.~\ref{sec:conjecture}. We discuss the scaling
behavior for the charge-$e/4$ non-Abelian quasihole in the Moore-Read
state in Sec.~\ref{sec:nonabelian} and generalize the discussion to
the Read-Rezayi states in Sec.~\ref{sec:readrezayi}. We summarize in
Sec.~\ref{sec:conclusion}.

\section{Model and earlier results}
\label{sec:model}

In the plane (disc) geometry we consider an FQH droplet at various filling
fractions, which correspond to the series of the Laughlin states, the
Moore-Read state, and the Read-Rezayi parafermion states. We generate
various Abelian and non-Abelian quasiparticles at the center of the
droplet. We assume a single-particle tunneling potential
\begin{equation}
V_{\rm tunnel}(\theta) = V_t \delta(\theta),
\end{equation}
which breaks the rotational symmetry. For the many-body states with
$N$ electrons, we write the tunneling operator as the sum of the
single-particle operators,
\begin{equation}
{\cal T} = \sum_{i=1}^N V_{\rm tunnel}(\theta_i) = V_t \sum_{i=1}^N
\label{eq:electrontunneling}
\delta(\theta_i).
\end{equation}
We compute the bulk-to-edge tunneling amplitude $\Gamma^{qh} =
\left \vert \langle \Psi_{\rm GS} \vert {\cal T} \vert \Psi_{\rm GS}^{qh}
\rangle \right \vert$, where $\Psi_{\rm GS}^{qh}$ and $\Psi_{\rm GS}$ are the FQH
ground states with and without a quasihole (at the disc center), respectively. For
convenience, we will henceforth set $V_t = 1$ as the unit of the tunneling
amplitudes. As seen in the earlier
work~\cite{chen09}, the matrix elements consist of contributions from
the respective Slater-determinant components $\vert l_1,...,l_N
\rangle \in \Psi_{\rm GS}$ and $\vert k_1,...,k_N \rangle \in
\Psi_{\rm GS}^{qh}$, where $l$s and $k$s are the angular momenta of the
occupied orbitals. A non-zero contribution only enters when $\vert
l_1,...,l_N \rangle$ and $\vert k_1,...,k_N \rangle$ are identical
except for a single pair $l_i$ and $k_j$ with the corresponding
angular momentum difference. More details are available in
Ref.~\cite{chen09}.

To be more relevant to the experimental situations in which
quasiparticles tunnel between two edges, we study the edge-to-edge
tunneling by inserting $n$ Laughlin quasiholes into the center of the
droplet~\cite{chen09}. This transforms a wavefunction $\Psi(\{z_i\})$
to $\prod_{i=1}^N z_i^n \Psi(\{z_i\})$, so that each component Slater
determinant becomes a new one, picking up a new normalization
factor. The first $n$ orbitals from the center are now completely
empty and the electrons are occupying orbitals above $n$, effectively 
producing an FQH droplet on an annulus. The tunneling distance $d(n,N)$ between
the inner and outer edges decreases monotonically under this
transformation. Correspondingly, $\Gamma^{qh}$ is defined as the
edge-to-edge tunneling amplitude.

The earlier work~\cite{chen09} found that the tunneling amplitude
ratio of quasiparticles with different charges decays with a Gaussian
tail as the interedge distance increases. The characteristic length
scale associated with this dependence originates partially from the
difference in the corresponding quasiparticle charges.  In the
Moore-Read state, for example, the tunneling amplitude for a charge
$e/4$ quasiparticle is larger than that for a charge $e/2$
quasiparticle~\cite{chen09,bishara09}. Our analyses~\cite{chen09} 
also show intriguing size dependence in the
tunneling amplitudes for the $e/4$ and $e/2$ quasiholes, although
their ratio appears to be size independent in the annulus
geometry. These observations motivated us to extend the study on the
size dependence of $\Gamma^{qh}$ for different types of quasiholes in
the Read-Rezayi series of FQH states, which include Laughlin and
Moore-Read states as special members.

We note that in Eq. (\ref{eq:electrontunneling}) we introduced the bare tunneling potential for
{\em electrons}, which form fractional quantum Hall liquids. Our results represent
the tunneling amplitudes for {\em quasiparticles} (not for electrons) and
have therefore taken into account the many-body correlations of the system. 
But for quasiparticles, when treated as elementary excitations of the system, these
are bare tunneling amplitudes at the microscopic length and energy
scales. They are subject to further renormalization when effective low-energy 
theories are constructed by integrating out degrees of freedom at higher-energy and shorter length scales.

\section{Field theoretical and numerical analyses of the tunneling
  amplitudes of Abelian quasiparticles}
\label{sec:fieldtheory}

We start with a field theoretical analysis of the quasiparticle
tunneling amplitude, which illustrates our calculation and provides an
expectation on the results. We consider, for illustration, a system of
electrons and quasiparticles on a cylinder with circumference $L$ and 
edge-to-edge distance $d \ll L$. This geometry is equivalent to an 
annulus with an
edge-to-edge distance much smaller than the radius. For fixed $d$, the
system size $N \propto L$. We assume that the edge runs around the $x$
direction, while tunneling occurs along the $y$ direction at $x=0$.

We introduce quasiparticle operators $\Psi_{a,j}(x)$, with $j=1,2$
corresponding to the two edges, while $a$ is quasiparticle type, and
normalize $\Psi_a$ (at each edge) such that the equal time Green's
function satisfies
\begin{equation}
\label{eq:qpgreen}
G_a(x-x')=\langle 0|\Psi_a^{\dagger}(x)\Psi_a(x')|0 \rangle
\sim |x-x'|^{-2\Delta_a},
\end{equation}
where $\Delta_a$ is the conformal dimension of $\Psi_a(x)$, and proper
factors of microscopic length scale $\ell$ are implied to ensure the
correct dimensionality of all quantities.

In a low-energy effective theory, the tunneling Hamiltonian,
transferring various types of quasiparticles from one edge to another
at $x=0$, takes the form
\begin{equation}
H_T = L \sum_a t_a [\Psi_{a,1}^\dagger(0)\Psi_{a,2}(0) + h.c.],
\end{equation}
where $t_a$ depends on quasiparticle type $a$ but has no $L$
dependence at fixed $d$. To facilitate comparison with numerical
calculations on rotationally invariant geometries, we include a
prefactor $L$---the Jacobian when transforming $\delta(\theta)$ on
the annulus to $\delta(x)$ on the cylinder.

A state generated by tunneling a quasiparticle from one edge to
another takes the following form (which is a momentum eigenstate):
\begin{equation}
\label{eq:quasiparticle}
\vert \Psi_a^{qh} \rangle = C_a \int_0^L{dxdx'}\Psi^\dagger_{a,1}(x)\Psi_{a,2}(x')|0\rangle.
\end{equation}
It is easy to show using Eq. (\ref{eq:qpgreen}) that the normalization
factor $C_a\propto L^{-2+2\Delta_a}$ for
\begin{equation}
\label{eq:relevance}
\Delta_a \le 1/2,
\end{equation}
in which case the corresponding quasiparticle tunneling operator is relevant in the renormalization group (RG) sense\cite{kane92b}.

We define the bare quasiparticle tunneling matrix element
\begin{eqnarray}
\label{eq:gamma}
\Gamma_a &=& \langle 0 | H_T | \Psi_a^{qh} \rangle \nonumber \\
&\propto& t_aL^{-1+2\Delta_a} \int{dxdx'}
\langle 0 | \Psi_{a,2}^\dagger(0)\Psi_{a,1}(0) \Psi^\dagger_{a,1}(x)\Psi_{a,2}(x')
| 0 \rangle \nonumber \\
&=& L^{1-2\Delta_a} K_a(d),
\end{eqnarray}
where we used the properties (\ref{eq:qpgreen}) and
(\ref{eq:relevance}) and $K_a(d)$ encodes $d$-dependence of $t_a$,
which is expected to be dominated by the Landau level gaussian
factor\cite{bishara09,chen09}. This scaling behavior is expected for
``elementary'' Abelian quasiholes of the Laughlin type, e.g., the
charge-$e/3$ quasiholes in the $\nu = 1/3$ Laughlin state, as well as
for the charge-$e/2$ quasihole (in the identity sector) in the $\nu =
1/2$ Moore-Read state.

\begin{figure}
\centering
\includegraphics[width=12cm]{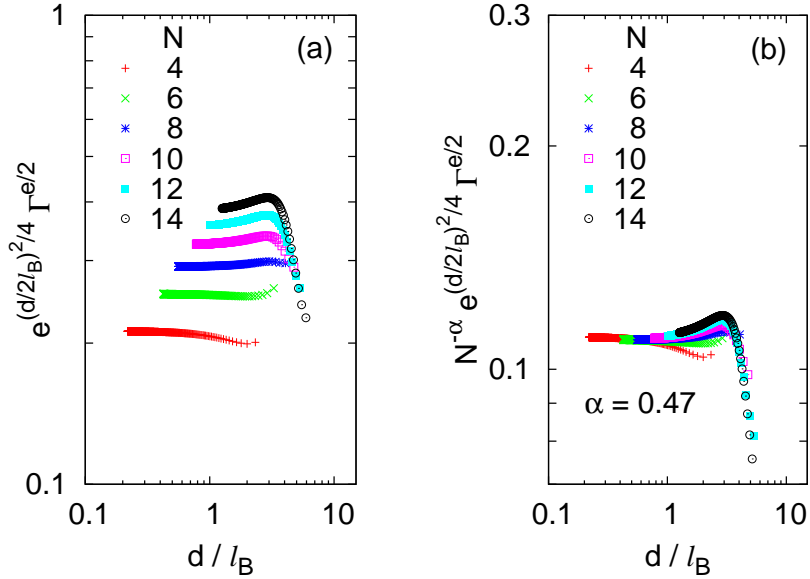}
\caption{\label{fig:half}(Color online) Rescaled tunneling amplitude
  (a) $e^{(d/4l_B)^2} \Gamma^{e/2}$ and (b) $N^{-\alpha}
  e^{(d/4l_B)^2} \Gamma^{e/2}$ with $\alpha = 0.47$ for the charge
  $e/2$ quasihole in the Moore-Read state as a function of the
  edge-to-edge distance $d$.  }
\end{figure}

\begin{figure}
\centering
\includegraphics[width=12cm]{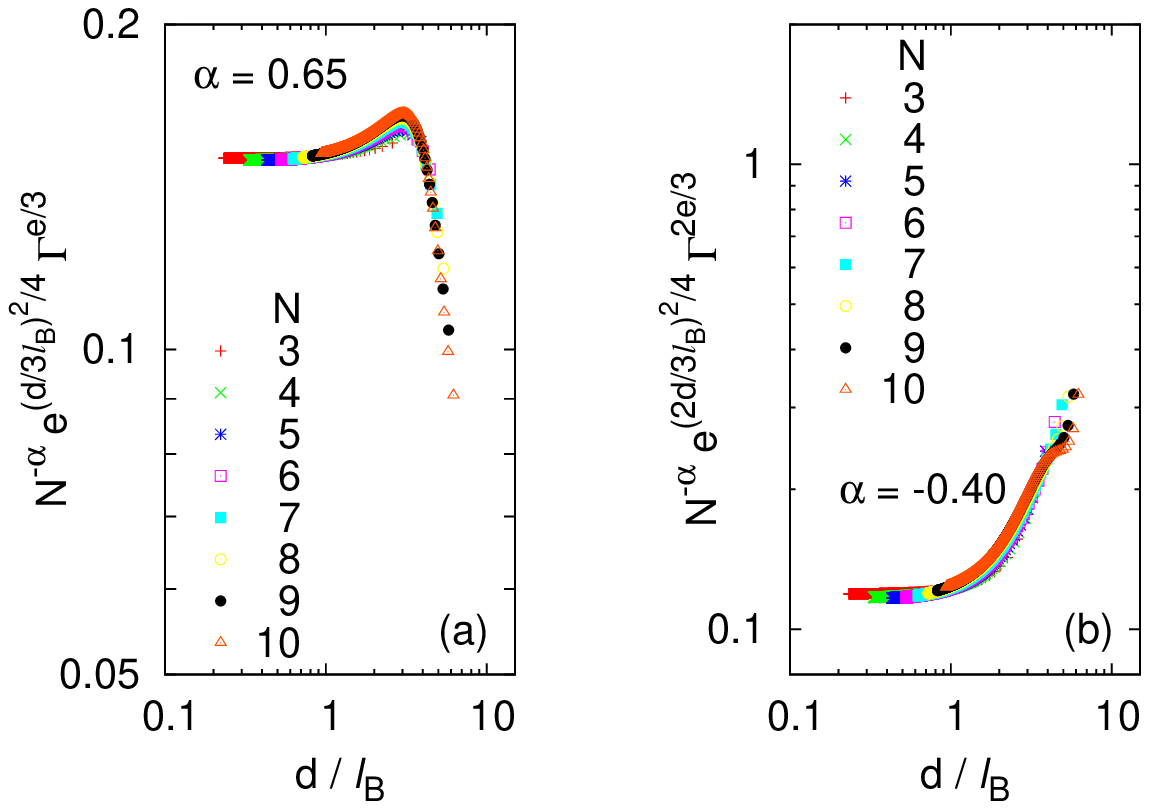}
\caption{\label{fig:third}(Color online) Rescaled tunneling amplitude
  $N^{-\alpha} e^{(qd/2el_B)^2} \Gamma^{q}$ for quasiparticles with
  (a) $q = e/3$, $\alpha = 0.65$ and (b) $q = 2e/3$, $\alpha = -0.4$
  in the Laughlin state at $\nu = 1/3$ as a function of the
  edge-to-edge distance $d$.  }
\end{figure}

We now compare the scaling behavior with numerical
results\cite{chen09}. For clarity, we multiply the tunneling
amplitude in Fig.~4(b) of Ref.~\cite{chen09} by a factor of
$e^{(d/4l_B)^2}$ ($l_B$ being the magnetic length) for the charge
$e/2$ quasihole in the Moore-Read state and plot the rescaled data in
Fig.~\ref{fig:half}(a). We find the rescaled data, depending on the
corresponding number of electrons $N$, falls on a series of
curves. Assuming the curves scale as $N^{\alpha}$, we obtain $\alpha =
0.47$ for the best scaling collapse, as shown in
Fig.~\ref{fig:half}(b). Similarly, we analyze and plot the
corresponding scaling collapses for charge $e/3$ and $2e/3$ quasiholes
in the Laughlin state at $\nu = 1/3$ in Figure~\ref{fig:third}.  We
obtain the optimal parameter $\alpha = 0.65$ and $-0.4$, respectively.  In
Table~\ref{tbl:exponent}, we compare the optimal fitting
$\alpha$ and the conformal dimensions $\Delta$ of the corresponding
quasiholes. We find excellent to reasonably good agreements with the relation
\begin{equation}
\alpha = 1 - 2\Delta
\label{eq:scalingbehavior}
\end{equation}
obtained above.  In the charge-$2e/3$ quasihole case for $\nu=1/3$, we
note that $\Delta = 2/3 > 1/2$ and, therefore, the condition of
Eq.~(\ref{eq:relevance}) is {\em not} satisfied. In addition, this is
a ``composite'' (instead of ``elementary'') quasihole, whose scaling
behavior requires a separate (and more complicated) analysis, which we
present below.

\begin{table}
  \caption{\label{tbl:exponent}
    The scaling exponent $\alpha$ of the quasihole tunneling amplitude
    and the corresponding conformal dimension of the quasiholes.}
\begin{center}
\begin{tabular}{c c c c}
\hline \hline
\hspace{0.5cm} $q$ ($\nu$) \hspace{0.25cm}
& \hspace{0.25cm} e/2 (1/2) \hspace{0.25cm}
& \hspace{0.25cm} e/3 (1/3) \hspace{0.25cm}
& \hspace{0.25cm} 2e/3 (1/3) \hspace{0.25cm} \\
\hline
$\Delta$ & 1/4 & 1/6 & 2/3 \\
$1 - 2 \Delta$ & 1/2 & 2/3 & -1/3 \\
$\alpha$ & 0.47 & 0.65 & -0.40 \\
\hline
\end{tabular}
\end{center}
\end{table}

The momentum eigenstate generated by tunneling a $2e/3$ quasihole from
one edge to another takes the form
\begin{equation}
\label{eq:2quasiparticles}
\vert \Psi_a^{2qh} \rangle = C_{2a} \int{dx_1dx_2dx'_1dx'_2}\Psi^\dagger_{a,1}(x_1)
\Psi^\dagger_{a,1}(x_2)\Psi_{a,2}(x'_1)\Psi_{a,2}(x'_2)|0\rangle,
\end{equation}
where $\Psi_a$ is the operator for an $e/3$ quasihole; the expression
above explicitly incorporates the fact that the $2e/3$ quasihole is a
composite object, and the state created by its tunneling moves two
$e/3$ quasiholes from one edge to another, which tunnel simultaneously
but are not necessarily bound together before and after the tunneling
process.

To calculate the normalization factor $C_{2a}$ and tunneling matrix
element $\langle 0 | H_T | \Psi_a^{2qh} \rangle$, we need the full
machinery of chiral Luttinger liquid theory for the $\nu=1/M$ Laughlin
state\cite{kane92b}, in which $\Psi_a(x)\sim \exp[i\varphi(x)/\sqrt{M}]$
and $\Psi_{2a}(x)\sim \exp[2i\varphi(x)/\sqrt{M}]$, where $\varphi$ is a
bosonic Gaussian field whose normalization is determined by the
conformal dimension of $\Psi_a$ which is $\Delta_a=1/2M$; we also have
$\Delta_{2a}=4\Delta_a$ following from the fact that $\varphi$ is a free or
Gaussian field. Using the chiral Luttinger liquid theory whose action
(for a single edge) takes the form\cite{kane92b}
\begin{equation}
S={M\over 4\pi}\int{dtdx}[(\partial_t+v\partial_x)\varphi(x,t)][\partial_x \varphi(x,t)],
\end{equation}
it is straightforward to calculate
\begin{equation}
C_{2a} \propto \left \vert \int{dx_1dx_1'dx_2dx_2'}
\langle 0 \vert e^{{i\over \sqrt{M}}[\varphi(x_1)+\varphi(x_2)-\varphi(x_1')-\varphi(x_2')]}
\vert 0 \rangle \right \vert^{-1} \propto
L^{-4+4\Delta_a}
\end{equation}
and
\begin{eqnarray}
\label{eq:gamma2}
\Gamma_{2a} &\equiv& \langle 0 | H_T | \Psi_a^{2qh} \rangle
\propto t_{2a}L^{-3+4\Delta_a} \left \vert \int{dxdx'}
\langle 0 | e^{{i\over \sqrt{M}}[2\varphi(0)-\varphi(x)-\varphi(x')] }
| 0 \rangle\right \vert^2 \nonumber \\
&=& L^{1-8\Delta_a} K_{2a}(d)=L^{1-2\Delta_{2a}} K_{2a}(d),
\end{eqnarray}
where we used the fact that $\Delta_{2a}=4\Delta_a$ in the last step.

Generalizing this analysis to tunneling of a charge $me/M$ quasiparticle
in Laughlin state at $\nu=1/M$, we find
\begin{eqnarray}
C_{ma}\propto
L^{-2m+2m\Delta_a}
\end{eqnarray}
and
\begin{eqnarray}
\Gamma_{ma}
=L^{1-2m^2\Delta_a} K_{ma}(d)=L^{1-2\Delta_{ma}} K_{ma}(d),
\end{eqnarray}
where we used the fact that $\Delta_{ma}=m^2\Delta_a$.  As
a result the relation (\ref{eq:scalingbehavior}) holds in all these
cases.

\section{Conjectures on exact amplitudes in a quasi-one-dimensional
  limit}
\label{sec:conjecture}

\subsection{The quasi-one-dimensional limit and the connection to Jack
  polynomials}

For the Laughlin state and the Moore-Read state, the numerical results
presented above agree with the scaling analyses, but not to a high
precision. For example, the exponent for the charge $2e/3$ quasihole
$\alpha = -0.4$ is 20\% smaller than the expectation value of
$-1/3$. Clearly, the systems are far from the thermodynamic
limit. This motivated us to study the scaling behavior from a
different approach: by conjecturing exact (or approximate) formulas
and extracting exact exponents from these conjectures. To achieve
that, we consider the quasi-one-dimensional $d \rightarrow 0$
limit~\cite{chen09}, in which the scaling behavior persists, as
indicated by Figs.~\ref{fig:half} and \ref{fig:third}.

In the mapping from disk to annulus we described earlier, the
wavefunctions, in terms of polynomials of electron coordinantes, are
unchanged; however the geometry, through the normalization of single-electron
basis, changes. We point out that in the $d \rightarrow 0$ limit,
there is no need to normalize each single-electron Landau level
orbital wavefunction by a momentum-dependent coefficient. When both
the inner and outer radii are much larger than their difference, the
normalization factor depends only on the number of quasiholes in the
lowest order, which is the same for all occupied orbitals. From a
different point of view, we can write down the antisymmetric many-body
ground state and quasihole wavefunctions as weighted sums of Slater
determinants $sl_{\mu} = \det \left (z_i^{\mu_j} \right )$. In the $d
\rightarrow 0$ limit, all Slater determinants are normalizable by the
same constant.\footnote{For a concrete example, the four-electron
  Moore-Read state in the $d \rightarrow 0$ limit, when we set $C =
  \sqrt{13 \cdot 5!  4!  3!  2!} / \sqrt{12}$ in Eq.~(C4) of
  Ref.~\cite{chen09}, contains exactly the same coefficients as the
  example in Ref.~\cite{bernevig09}.}  As a result, the insertion of an
additional Abelian quasihole only changes the labels of the orbitals
without affecting the amplitude of individual Slater determinants and the
overall normalization factor.

With the recent development of the
connection~\cite{haldane06,bernevig08} of Jack
polynomials~\cite{stanley89} with a negative Jack parameter
$\alpha_J$ and fractional quantum Hall wavefunctions, we now
understand that these antisymmetric quantum Hall wavefunctions can be
written as single Jack polynomials multiplied by the Vandemonde
determinant (which are sums of Slater determinants) whose
corresponding amplitudes can be evaluated
recursively~\cite{bernevig09}. We emphasize that the amplitudes are
integers up to a global normalization constant $1 / \sqrt{C}$, where
$C$ is an integer. The Jack polynomial connection facilitates {\it the
  exact evaluation of the tunneling amplitude} even in relatively
large systems. Otherwise, one would need Lanczos diagonalization to
produce a numerical approximation with an accuracy that depends on the
number of iterations, which is only cost effective for sparse
Hamiltonians. For multiparticle interactions the Hamiltonian becomes
very dense and the Lanczos algorithm becomes progressively more
expensive.  Based on the exact results, we can conjecture~\cite{lee10}
the functional forms of the scaling functions for the Laughlin states,
the Moore-Read state, and the Read-Rezayi ${\mathbb Z}_k$ parafermion
states.

\subsection{Scaling of quasihole tunneling amplitudes in the Laughlin
  states}

The Laughlin wavefunction at filling fraction $\nu = 1/M$
can be constructed by the chiral boson conformal field theory (CFT) 
with a compactification
radius $M$~\cite{moore91}. The primary fields are vertex operators
$e^{i m \varphi(z) / \sqrt{M}}$, where $\varphi(z)$ is the chiral
boson. Operators with $m = 1, 2, \dots M$ correspond to quasiholes ($m
< M$) or electrons ($m = M$), whose conformal dimensions are
$\Delta(m,M) = m^2/(2M)$.

For $M = 3$ or $\nu = 1/3$, we conjecture the tunneling amplitude for
the charge-$e/3$ quasihole is
\begin{equation}
\label{eq:laughlinScaling}
2 \pi \Gamma^{e/M}_{L, M} (N) = {N \over M}
B \left (N, {1 \over M} \right ),
\end{equation}
where $M = 3$ and $N$ is the number of electrons. Here we introduce
the beta function $B(x, \beta) = \Gamma(x) \Gamma(\beta) /
\Gamma(x+\beta)$ which, for large $x$ and fixed $\beta$, asymptotically 
approaches $\Gamma(\beta) x^{- \beta}$, where $\Gamma(x)$ is the Gamma
function (not the tunneling amplitude elsewhere). We verified
numerically that the conjecture is {\it exact for up to 10 electrons};
therefore, assuming the conjecture is also exact for larger system, we
obtain the exact scaling exponent $\alpha^{e/3} = 1 - 1/3 = 2/3$.
This is also verified to be correct for $M =
5$.\footnote{Eq.~(\ref{eq:laughlinScaling}) also applies to the
  integer case ($M = 1$), in which the righthand side reduces to
  unity.} In other words, based on the scaling analysis we discussed
earlier, we can compute the conformal dimension of charged Abelian
quasiholes in the Laughlin state to be $\Delta^{1}_{M} = 1 / (2M)$.

Interestingly, we can make another connection to Jack polynomials by
rewriting the tunneling amplitude in a neat way as, e.g. for $\nu = 1/3$,
\begin{equation}
\label{eq:rootform}
2 \pi \Gamma^{e/3}_{L, M = 3} (N) = N
{{\hat \Omega (10010010...01001)\,\,\, } \over
{\hat \Omega  (01001001...001001)}},
\end{equation}
where the operator $\hat \Omega$ takes the product of the occupied
nonzero single-particle momenta, e.g., $\hat \Omega (10010010...01001)
= 3 \cdot 6 \cdot \cdots \cdot (3N-3) = (3N-3)!!!$. One recognizes that
the arguments of $\hat \Omega$ are precisely the root configurations of
the corresponding Laughlin ground state and the charge-$e/3$ quasihole
state, which are the final and initial states, respectively, of the
quasihole tunneling process. 

The exact tunneling amplitude for charge-$2e/3$ quasiholes in the
Laughlin state discussed earlier can be written as
\begin{equation}
\label{eq:rootform2over3}
2 \pi \Gamma^{2e/3}_{L, M = 3} (N) = 2! N
{{\hat \Omega (101101...011)\,\,\,\,\,\, } \over
{\hat \Omega  (011011...11011)}}
= 2! N {{\hat \Omega \left (
\begin{array}{l}
001001...01 \\
100100...001
\end{array}
\right )\,\,\,\,\,\, } \over
{\hat \Omega  \left (
\begin{array}{l}
010010...1001 \\
001001...01001
\end{array}
\right )}},
\end{equation}
where $\hat \Omega \left ( \,^\lambda_\mu \right ) = \hat \Omega
(\lambda) \hat \Omega (\mu)$. The first equality can be understood as
the particle-hole transformation of the charge-$e/3$ quasihole
tunneling amplitude, implying the tunneling of a $2e/3$ quasihole is
equivalent to the tunneling of a $e/3$ quasiparticle. Formally, the
second equality can be understood as decomposing the $2e/3$ quasihole
into two charge $e/3$ quasiholes. By studying $\Gamma^{2e/3}_{L,3}
(N+1) / \Gamma^{2e/3}_{L,3} (N)$, we conclude that the scaling
behavior of $\Gamma^{2e/3}_{L,3} (N) \sim N^{-1/3}$ is again
consistent with Eq. (\ref{eq:scalingbehavior}) for
\begin{equation}
\label{abelianDimension}
\Delta^{me/M}_{L, M} = {m^2 \over 2M}
\end{equation}
as expected. We note that without the exact amplitude conjecture we
would obtain a large (20\%) error of the exponent based on finite-size
scaling only; this means that the systematic error due to
finite-system size is not negligible unless we can conjecture
numerically exact results.

We can write down similar results for the $\nu = 1/5$ Laughlin state,
which are in agreement with Eq.~(\ref{abelianDimension}) with $M = 5$
for $m = 1$-4. For example, for $m = 3$,
\begin{equation}
\label{eq:rootform3over5}
2 \pi \Gamma^{3e/5}_{L, 5} (N) = 3! N
{{\hat \Omega \left (
\begin{array}{l}
0001000010...001 \\
0000100001...0001 \\
1000010000...00001
\end{array}
\right )\ \ \ \ \  } \over
{\hat \Omega  \left (
\begin{array}{l}
0100001000...100001 \\
0010000100...0100001 \\
0001000010...00100001
\end{array}
\right )}}.
\end{equation}
The scaling behavior is asymptotically $\Gamma^{3e/5}_{L, 5} \sim N^{-4/5}$, again consistent with Eq. (\ref{eq:scalingbehavior}).

\subsection{Scaling conjecture for Abelian charge-$e/2$ quasiholes in
  the Moore-Read state}

The Moore-Read wavefunction at filling fraction $\nu = 1/2$ can be
constructed by the Ising CFT, which describes the neutral fermion
component, and the chiral boson CFT, which describes the charge
component~\cite{moore91}. Two quasihole operators relevant to
interedge tunneling are $\psi_{qh}^{e/4} = \sigma e^{i \varphi / 2
  \sqrt{2}}$ and $\psi_{qh}^{e/2} = e^{i \varphi / \sqrt{2}}$. The
former is a non-Abelian quasiparticle, while the latter an Abelian
one. We note that the charge-$e/2$ quasihole can be regarded as one of
the two fusion results (i.e., $\sigma \times \sigma = 1 + \psi$) of
two charge-$e/4$ quasiholes; the other, $\psi_{qh}^{e/2,\psi} = \psi
e^{i \varphi / \sqrt{2}}$, is irrelevant (in the RG sense) in interedge tunneling. 
The conformal dimensions of the charge $e/2$ quasihole is $\Delta^{e/2} = 1/4$.

We find the tunneling amplitude for $\psi_{qh}^{e/2}$ in the $d
\rightarrow 0$ limit to be {\it exactly}
\begin{equation}
2 \pi \Gamma^{e/2}_{MR} (N) = N
{{\hat \Omega (11001100110...0110011)} \over
{\hat \Omega (011001100110...0110011)}}.
\end{equation}
This is similar to Eq.~(\ref{eq:rootform}) for the $\nu = 1/3$ Laughlin
case, emphasizing, again, the role of root configuration of the states
involved in the tunneling process. One can write, equivalently,
\begin{equation}
\label{eq:MRScaling}
2 \pi \Gamma^{e/2}_{MR} (N) = {N \over 4}
B \left ({N \over 2}, {1 \over 2} \right ),
\end{equation}
which leads to $\Gamma^{e/2}_{MR} (N) \sim N^{1/2}$, again
consistent with the scaling analysis, i.e. $\alpha^{e/2} = 1 - 2
\Delta^{e/2}$.

\section{Scaling analysis for non-Abelian quasiholes
in the Moore-Read state}
\label{sec:nonabelian}

We have seen in the previous two sections that the scaling behavior of
the Abelian quasihole tunneling amplitudes can be well understood. The
individual scaling exponent is simply related to the conformal
dimension of the tunneling particle. In this section, we focus on the
non-Abelian charge-$e/4$ quasihole in the Moore-Read phase. The
quasihole operator can be written as $\Psi^{e/4}_{qh} = \sigma e^{i
  \varphi / 2\sqrt{2}}$, which consists of a bosonic charge component
with conformal dimension $\Delta^{e/4}_c = 1/16$ and a fermionic
neutral component also with conformal dimension $\Delta^{e/4}_n =
1/16$. The total dimension is thus $\Delta^{e/4} = \Delta^{e/4}_c +
\Delta^{e/4}_n = 1/8$.  In some sense, the situation for the
charge-$e/4$ quasihole in the Moore-Read state is somewhat similar,
but not identical to the $2e/3$ quasiparticle at $\nu=1/3$, as it
carries a charge component and neutral component. It is thus a ``composite''
object.

Incorporating our prior knowledge of the Abelian cases, we carefully
analyze the tunneling amplitude of the non-Abelian quasihole in the
quasi-one-dimensional limit and conjecture that for the charge $q = e/4$
quasihole in the Moore-Read state with $N = 2n$ electrons, the
tunneling amplitude is
\begin{equation}
\label{eq:nonabelian}
2 \pi \Gamma^{e/4} (N) = {N/2 \over 4}
  \sqrt{ B \left ({N \over 2}, {1 \over 2} + {\sqrt{3} \over 4} \right )
    B \left ({N \over 2}, {1 \over 2} - {\sqrt{3} \over 4} \right )}.
\end{equation}
The square-root form,which is absent in the Abelian cases, was conceived 
by noting that the ground state and the state with quasi-holes 
differ because of presence of twists ($\sigma$s at the center and along the
edge) in the latter.  Therefore, the two wavefunction normalization constants
(square roots of inverse integers) are not equal and the square root
does not disappear from the tunneling amplitude. The second arguments of the two
Beta functions turn out to be the solutions of $x^2 - x + 1/16 =0$. We
emphasize that the formula is verified to be {\it exact to the machine
precision  ($< 10^{-15}$)for up to 18 electrons}. This implies that it
has the same scaling behavior $\Gamma^{e/4}_{MR} (N) \sim N^{1/2}$ as
that of the Abelian charge-$e/2$ quasiholes.

\begin{figure}
\centering
\includegraphics[width=12cm]{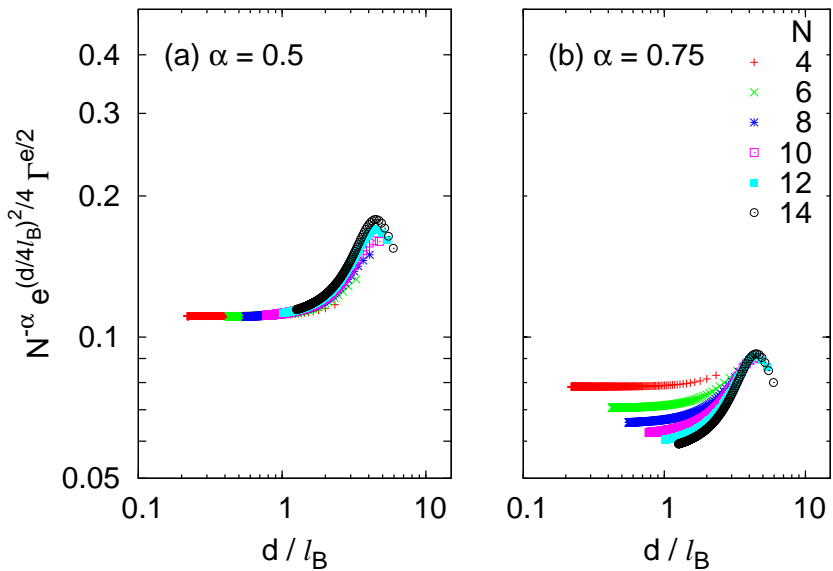}
\caption{\label{fig:quarter}(Color online) Rescaled tunneling
  amplitude $N^{-\alpha} e^{(d/4l_B)^2/4} \Gamma^{e/4}$ for charge
  $e/4$ quasiholes in the Moore-Read state as a function of the
  edge-to-edge distance $d$ for (a) $\alpha = 0.5$ and (b) $\alpha =
  0.75$. }
\end{figure}

This result is very different from those of the Abelian quasiholes.
Clearly, the scaling exponent $\alpha \neq 1 - 2 \Delta^{e/4} = 3/4$,
as expected from simple dimension counting. We check the reduced
tunneling amplitudes at finite edge-to-edge distance $d$ and compare
the scaling collapses with $\alpha = 0.5$ and $\alpha = 1 -
2\Delta^{e/4} = 0.75$ in Fig.~\ref{fig:quarter}. We find that the
choice of $\alpha = 0.5$ yields a much better scaling collapse
especially for $d < 3l_B$.

While we do not have a satisfactory theory to explain the anomalous
scaling behavior for the non-Abelian quasihole, we speculate that one of
the potential explanations may be as follows. In the quasi-one-dimensional
limit, the two edges may not be regarded as independent edges for the
neutral component. It is likely that we need to include coupling
between neutral components on the two edges (the Abelian charge
components are not affected). If the coupling is relevant,
we can estimate the length scale for such interaction to be $\sim
3l_B$, which is in agreement with the earlier
estimate~\cite{baraban09}. Beyond this scale topological ground state
degeneracy and unitary transformation due to braiding are
exponentially exact. However, this argument cannot explain why the
exponent happens to be $1/2$.

Alternatively, one may speculate that the charge and neutral components may
not be always bound together. A realistic tunneling potential, often
arising from applying a gate voltage, couples only to the charge
component giving neutral components freedom
to propagate in the bulk region other than $x = 0$.  Qualitatively, we
expect the scaling behavior will be different from simply replacing
$\Delta^{e/4}$ with the sum of the charge and neutral conformal
dimensions, $\Delta^{e/4}_c + \Delta^{e/4}_n$ in
Eq.~(\ref{eq:scalingbehavior}).  In general, the tunneling process may
allow additional $\sigma$-propagators, which may help produce the
exponent $\alpha = 1/2$ as $\alpha = 1 - 2 \Delta^{e/4}_c - 6
\Delta^{e/4}_n$ with an anomalous exponent $\delta \alpha = -4
\Delta^{e/4}_n$.

\section{Speculations on the Read-Rezayi $\mathbb{Z}_k$
  parafermion states}
\label{sec:readrezayi}

To offer additional insight, we attempt to generalize the results to
the Read-Rezayi $\mathbb{Z}_k$ parafermion states with the electron
operator
\begin{equation}
\psi_{e} = \psi_1 e^{i \sqrt{{k+2} \over k} \varphi}.
\end{equation}
The conformal dimension for $\psi_1$ is ${{k-1} \over k}$, while for
the vertex operator it is ${{k+2} \over {2k}}$. The filling fraction
is $\nu_k = {k \over {k+2}}$.  In practice, we generate this ground
state by a Jack parameter $\alpha_J = -(k+1)$ and the corresponding root
configuration of $1^k001^k00 \cdots 1^k$ (where $1^k$ means $k$
consecutive 1s) so that there are exactly $k$ 1s in any $(k+2)$
consecutive orbitals.

The charge ${e \over {k+2}}$ non-Abelian
quasihole operator is
\begin{equation}
\psi_{qh}^{e/(k+2)} = \sigma_1 e^{{i\varphi} \over \sqrt{k(k+2)}}.
\end{equation}
The conformal dimension for $\sigma_1$ is $\Delta_n = {{k-1} \over
  {2k(k+2)}}$ and for the vertex operator it is $\Delta_c = {1 \over
  {2k(k+2)}}$. One can form an Abelian quasihole of charge ${{ke}
  \over {k+2}}$ by fusing $k$ $\psi_{qh}^{e/(k+2)}$ quasiholes. The
conformal dimension of the Abelian quasihole is $\Delta_{ke/(k+2)} =
{k \over {2(k+2)}}$. The corresponding root configurations for the
smallest-charged non-Abelian and Abelian quasiholes are
$1^{k-1}0101^{k-1}010 \cdots 1^{k-1}01$ and $01^k001^k00 \cdots 1^k$,
respectively. The $e/4$ and $e/2$ quasiholes in the Moore-Read states
correspond to the $k = 2$ cases.

From Eqs.~(\ref{eq:laughlinScaling}) and (\ref{eq:MRScaling}), we
conjecture that the tunneling amplitude for the charge-${{ke} \over {k+2}}$
Abelian quasihole in the filling factor $\nu = {k \over {k+2}}$ state is
\begin{equation}
\label{eq:abelianScaling}
2 \pi \Gamma^{ke/(k+2),1}_{k} (N) = {N \over {k+2} }
B \left (N, {k \over {k+2}} \right ).
\end{equation}
We compare with the numerical results based on the recursive
construction and find that Eq.~(\ref{eq:abelianScaling}) is {\it not
  exact}, but the errors for states ($M = 1$) up to $k = 5$ are {\it
  all within 1\%}. This leads to
\begin{equation}
\Gamma^{ke/(k+2),1}_{k} (N) \sim N^{1-{k \over {k+2}}} \equiv N^{1-2\Delta_{ke/(k+2)}},
\end{equation}
which implies $\Delta_c \equiv \Delta_{e/(k+2)} = {1 \over
  {2k(k+2)}}$.

\begin{table}
  \caption{\label{tbl:abelianRR}
    The tunneling amplitude for charge-$ke/(k+2)$ Abelian quasiholes
in the Read-Rezayi states. They are all within 1\% error of
Eq.~(\ref{eq:abelianScaling}).}
\begin{center}
\begin{tabular}{c c c c}
\hline \hline
\hspace{0.5cm} $N/k$ \hspace{0.25cm}
& \hspace{0.25cm} k = 3 \hspace{0.25cm}
& \hspace{0.25cm} k = 4 \hspace{0.25cm}
& \hspace{0.25cm} k = 5 \hspace{0.25cm} \\
\hline
2 & 1.256203474 & 1.206153846 & 1.171688187 \\
3 & 1.451788763 & 1.358816509 & 1.296273516 \\
4 & 1.614288884 & 1.483200501 & 1.396827446 \\
5 & 1.755379103 & 1.589612764 & 1.481715173 \\
6 & 1.881240395 & 1.683409192 & 1.555472123 \\
7 & 1.995594026 &  &  \\
\hline
\end{tabular}
\end{center}
\end{table}

\begin{figure}
\centering
\includegraphics[width=12cm]{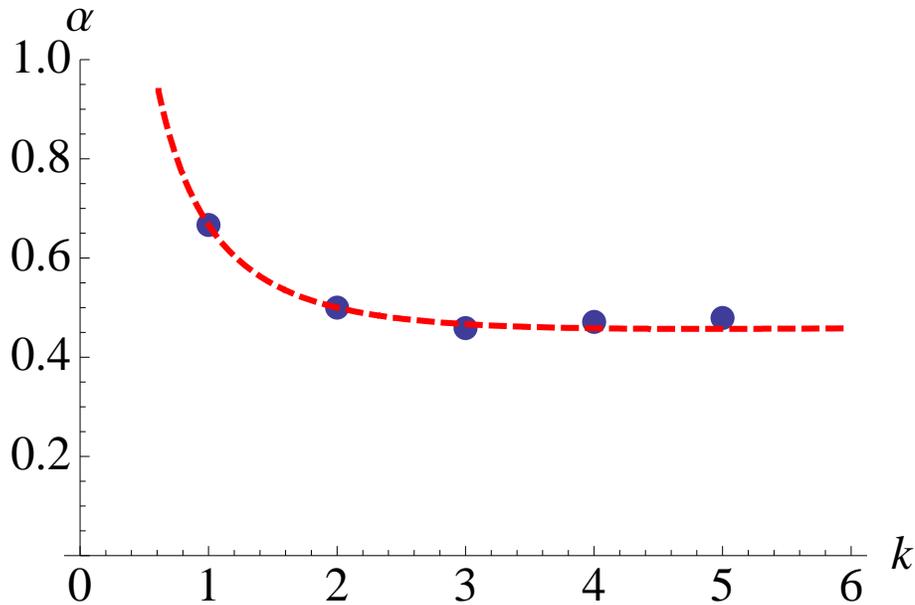}
\caption{\label{fig:fit}(Color online) Scaling exponent $\alpha$ for
  the smallest-charge quasihole tunneling amplitude [$\Gamma(N) \sim
  N^{\alpha}$] for the Read-Rezayi series with $k = 1$-5. The dashed
  line attempts to fit the exponent to a linear dependence on the
  conformal dimensions of the charge and neutral components
  [Eq.~(\ref{eq:fit})].}
\end{figure}

We want to obtain a similar approximation for the charge-$e/(k+2)$
non-Abelian quasihole, so that we can compute the conformal dimension
of $\sigma_1$. Ideally, the form should reduce to
Eq.~(\ref{eq:nonabelian}) for $k = 2$ and
Eq.~(\ref{eq:laughlinScaling}) for $k = 1$ (i.e., $M = 3$). But with
the origin of the numerous parameters in Eq.~(\ref{eq:nonabelian})
unclear, the attempt has not yet been successful. Instead, we fit the
numerical results to a power law in each case and list the exponents
in Table~\ref{tbl:readrezayi}, in addition to the case of $k = 1$ and
2 for the Read-Rezayi series. In Fig.~\ref{fig:fit}, we attempt to fit
the exponent to the form $\alpha^{e/(k+2)} = 1 - (sk+t) \Delta_c - (uk+v)
\Delta_n$, where $s$, $t$, $u$ and $v$ are integers. The linear
$k$-dependence in the fitting form takes into account the clustering
nature of the Read-Rezayi states. The result with the best fit is
\begin{equation}
\label{eq:fit}
\alpha^{e/(k+2)} = 1 - {{k^2 + 3k -2} \over {2k(k+2)}},
\end{equation}
as indicated by the dashed line in Fig.~\ref{fig:fit}.
Interestingly,
\begin{equation}
\alpha^{e/(k+2)} = 1 - 2 \Delta_c - 2 \Delta_n - {{k - 1} \over {2k}}.
\end{equation}
Incidentally, the last term (or the anomalous exponent) is
$-(k+2)\Delta_n$.

\begin{table}
  \caption{\label{tbl:readrezayi} The scaling exponent $\alpha$ for
    the smallest charge-$e/(k+2)$ quasihole tunneling amplitude for the
    Read-Rezayi series. They are obtained from the exact conjectures
    (for $k = 1$-2) or by fitting data in Table.~\ref{tbl:abelianRR}
    (for $k = 3$-5).}
\begin{center}
\begin{tabular}{c c c c c c}
\hline \hline
\hspace{0.25cm} $k$ \hspace{0.25cm}
& \hspace{0.25cm} 1 \hspace{0.25cm}
& \hspace{0.25cm} 2 \hspace{0.25cm}
& \hspace{0.25cm} 3 \hspace{0.25cm}
& \hspace{0.25cm} 4 \hspace{0.25cm}
& \hspace{0.25cm} 5 \hspace{0.25cm} \\
\hline
$\alpha$ & 2/3 & 1/2 & 0.4586 & 0.4711 & 0.4792 \\
\hline
\end{tabular}
\end{center}
\end{table}

\section{Summary and discussion}
\label{sec:conclusion}

In summary, we find that the tunneling amplitude for Abelian
quasiparticles exhibits finite-size scaling behavior with an exponent
related to the conformal dimension of the quasiparticles, irrespective of
whether their inter-edge tunneling is relevant or not.  This is true
for Abelian quasiparticles in both Abelian and non-Abelian quantum
Hall states.  Generically, we find that in our model the inter-edge
tunneling amplitude for an ideal quasiparticle (arising from the
variational wavefunctions) with charge $q$ and a conformal dimension
of $\Delta^q$ can be expressed as
\begin{equation}
\Gamma^q (N, d) = \Gamma_0 N^{\alpha^q} e^{-(qd/2el_B)^2},
\end{equation}
where $\alpha^q = 1 - 2 \Delta^{q}$ for an Abelian quasiparticle with
charge component only, e.g., $\alpha^{e/2} = 1/2$ for the charge $e/2$
Abelian quasihole in the Moore-Read state.  We note that $\Gamma_0$ is
related to the propagation of charge bosons and neutral (para)fermions
perpendicular to the edges, which contain additional dependency on $d$
as observed for $d > l_B$.  The observation of the scaling
behavior suggests that the systems are described by underlying
conformal field theories; in fact, the conformal dimensions of the
Abelian quasiholes obtained from the tunneling amplitudes are in
perfect agreement with those in the $\mathbb{Z}_k$ parafermion
theories for quantum Hall wavefunctions, based on which we can deduce
the conformal dimensions of non-Abelian quasiholes. Computing the
conformal dimensions of quasiparticles from wavefunctions has also
been attempted in the pattern of zeros classification~\cite{wen08} and
in the Jack polynomial approach~\cite{bernevig09b}.

The scaling behavior can be alternatively expressed by a differential
equation
\begin{equation}
\label{eq:rg}
\frac{\partial \tilde{\Gamma}^q}{\partial l} = \alpha^q
\tilde{\Gamma}^q = (1 - 2 \Delta^q)
\tilde{\Gamma}^q,
\end{equation}
where $\tilde{\Gamma}^q = e^{(qd/2el_B)^2} \Gamma^{q}$ and $N =
e^l$. Here we fix the edge-to-edge distance $d$ and the filling
fraction $\nu$ so the number of electrons $N \sim Ld$, where $L$ is the
length of the edge; in the large $N$ limit the annulus is thin so we
do not need to distinguish the lengths of the inner and outer edges.
We note that Eq.~(\ref{eq:rg}) resembles the renormalization group
flow equation in the context of edge state transport~\cite{kane92b}.
In particular, $\alpha^{2e/3}$ for the quasiparticles with charge
$2e/3$ is negative, which reflects that the quasiparticles are
irrelevant to inter-edge tunneling.

For the charge-$e/4$ non-Abelian quasihole in the Moore-Read state, we
find $\alpha^{e/4} = 1/2$ (not 3/4) and we speculate that the
contributions from the charge and neutral components are asymmetric.
Interestingly, the scaling exponent coincides with that of the
charge-$e/2$ Abelian quasiparticle and therefore we obtain perfect
data collapse in Fig.~5 of Ref.~\cite{chen09} for different
$N$. Generically, in the non-Abelian quasiparticle tunneling
amplitudes for the Read-Rezayi $\mathbb{Z}_k$ parafermion states, we
find anomalous scaling behavior (hence the signature of non-Abelian
statistics in model simulations) beyond simple scaling analysis.

X.W. thanks Dimitry Polyakov, Steve Simon, Smitha Vishveshwara, and
Zhenghan Wang for stimulating discussions. This work was supported by
DOE grant No. DE-SC0002140 (Z.X.H., E.H.R. and K.Y.) and the 973
Program under Project No. 2009CB929100 (X.W.). Z.X.H., K.H.L, and
X.W. acknowledge the support at the Asia Pacific Center for
Theoretical Physics from the Max Planck Society and the Korea Ministry
of Education, Science and Technology.

\end{document}